\documentclass[%
]{algotel}
\usepackage[utf8]{inputenc}
\usepackage[francais]{babel}
\usepackage{etex}
\usepackage[T1]{fontenc}

\usepackage{graphicx}
\usepackage[rightcaption]{sidecap}

\usepackage{pifont}
\PassOptionsToPackage{table,xcdraw}{xcolor}
\usepackage{adjustbox}
\usepackage{array}
\usepackage{multirow}
\usepackage{tikz}
\usetikzlibrary{decorations.pathreplacing}
\usepackage{booktabs}
\newcolumntype{R}[2]{%
	>{\adjustbox{angle=#1,lap=\width-(#2)}\bgroup}%
	l%
	<{\egroup}%
}
\newcommand*\rot{\multicolumn{1}{R{90}{1em}}}

\usepackage[printonlyused]{acronym}
\acrodef{WebRTC}{\emph{Web Real-Time Communication}}
\acrodef{TGDH}{\emph{Tree Group-Diffie Hellman}}
\acrodef{ICN}{\emph{Information-Centric Networks}}
\acrodef{VoICN}{\emph{Voice-over-ICN}}
\acrodef{BE}{\emph{Broadcast Encryption}}
\acrodef{IBE}{\emph{Identity-Based Encryption}}
\acrodef{ABE}{\emph{Attribute-Based Encryption}}
\acrodef{CDH}{\emph{Computational Diffie-Hellman}}
\acrodef{DDH}{\emph{Decisional Diffie-Hellman}}
\acrodef{SRTP}{\emph{Secure Real-time Transport Protocol}}
\acrodef{CDN}{\emph{Content Delivery Network}}
\acrodef{TLS}{\emph{Transport Layer Security}}
\acrodef{DTLS}{\emph{Datagram Transport Layer Security}}
\acrodef{ECC}{\emph{Elliptic Curve Cryptography}}
\acrodef{TCP}{\emph{Transmission Control Protocol}}
\acrodef{mKEM}{\emph{Multikey Encapsulation Mechanism}}
\acrodef{KGS}{\emph{Key Generation Center}}
\acrodef{HMAC}{\emph{keyed-Hash for Message Authentication Code}}
\acrodef{TGDH}{\emph{Tree-Based Group Diffie-Hellman}}
\acrodef{STR}{\emph{Skinny Tree Diffie-Hellman}}
\acrodef{PFS}{\emph{Perfect Forward Secrecy}}

\newcommand{\figref}[1]{figure~\ref{fig:#1}}

\newcommand{\tabref}[1]{table~\ref{tab:#1}, page \pageref{tab:#1}}

\author{Pierre-Antoine Rault\addressmark{1}\addressmark{3}, Luigi Iannone\addressmark{1}\addressmark{2}\thanks{Ce travail de recherche a été financé dans le cadre de la chaire NewNet@Paris de Télécom ParisTech.}{\ }}

\title[Group Key Agreement with TGDH in ICN]{Group Key Agreement in Information Centric Networks with Tree Group Diffie-Hellman}

\address{\addressmark{1}Télécom ParisTech, LINCS, 23 avenue d'Italie,
Paris, France\\\addressmark{2}prénom.nom@telecom-paristech.fr | \addressmark{3}sendmemail+tpt@rigelk.eu \\}


\begin{document}
\maketitle

\begin{abstract}
The client-server model is known to scale badly without redundant servers and caches. Information-Centric Networks (ICN) are an alternative designed to lessen that problem by distributing load to intermediary routers, lowering the bar to techniques only accessible to server farms. However, this seamlessness comes with its own set of problems. Communication security, by using a key per connection, prevents intermediaries from reading and thus caching content. We thus need to share a common key for larger groups of correspondents to be able to benefit from the advantages of ICN.
This paper presents the algorithmic choices to establish such a group key, the adaptations made for ICN, its implementation, and the statistics backing its performances comparison with our baseline. The first results tend to show a performance improvement over the selected base algorithm.
\end{abstract}

\section{Introduction}
\label{sec:in}

Les réseaux modernes doivent faire face à la massification des échanges sur Internet, et les dernières décennies n'ont pas ou peu changé la distribution du contenu,  qui y reste majoritairement centralisée. Si le paradigme client-serveur est simple à penser et mettre en place, des approches adressant \emph{du contenu} au lieu de \emph{machines où trouver du contenu} voient le jour à différents niveaux : par l'adressage de contenu sur une table de hachage distribuée, ou plus récemment par la mise en cache de contenu au niveau du réseau lui-même avec les réseaux centrés sur l'information (ICN). Ces derniers prennent en compte les capacités des nœuds intermédiaires à stocker de l'information, combinée à une gestion légère du cache, pour stocker tout paquet en vue de répondre à une répétition de sa demande dans un court laps de temps.

Les réseaux ICN étant des intermédiaires, ils voient passer du contenu de plus en plus souvent chiffré de bout en bout entre consommateur et producteur dudit contenu. Cela implique qu'un même contenu sera chiffré autant de fois différemment qu'il est demandé par des consommateurs différents, empêchant sa mise en cache efficace par réseau. Établir une clé commune pour un groupe de clients demandant le contenu à un même moment permettrait de rétablir la capacité de mise en cache d'ICN.

Au-delà des réseaux ICN, la problématique d'efficacité du cache d'un système de partage de contenu peut être appliquée à d'autres systèmes de communication dès lors qu'ils peuvent voir un bénéfice de l'utilisation de multicast dans leur modèle de communication. Cela concerne ainsi les couches applicatives comme WebRTC, qui permet le partage direct de contenu entre deux navigateurs et chiffre là encore avec une clé propre à la connection. Étant typiquement utilisé pour des vidéoconférences, WebRTC pourrait notamment bénéficier de l'utilisation d'un clé commune aux membres d'une vidéoconférence.

Pour améliorer la performance du cache dans ces deux cas d'usage, on effectue une sélection d'algorithmes existants d'établissement de clé de groupe et, après avoir définit des critères simples de comparaison, nous en choisissons un que nous implémentons. Après analyse de ses performances, nous proposons des améliorations potentielles dans le cadre de son utilisation pour ICN.

\section{Sélection des protocoles}
\label{sec:first}

Le domaine des protocoles d'établissement de clé entre plusieurs pairs est riche et répond à divers besoins et équilibres entre complexité communicationnelle et computationnelle. Certains de ces protocoles affichent des propriétés plus intéressantes pour les ICN que d'autres (voir table \ref{tab:intro:new_proto_selection}). La capacité d'\textbf{établir sans échanger} de clé, où deux parties ou plus s'accordent sur une clé commune sans transférer de clé et en s'assurant que des tiers ne puissent pas influencer la génération de leur clé. Cela permet de ne pas se reposer sur un nœud central de génération, stockage et transfert de clé. La \textbf{résistance aux partitions} de réseau, qui arrivent plus couramment à mesure que le groupe croît en taille. La \textbf{facilité d'implémentation}, qui est probablement le critère le plus subjectif mais un point important d'inquiétude alors que peu des algorithmes présentés sont open source et que certains concepts utilisés ont été peu étudiés. Le \textbf{nombre de tours} ou requêtes à chaque membre du groupe pour modifier celui-ci est une métrique impactant la performance d'autant plus que le délai réseau peut être important, même si l'on note que la notion de tour varie légèrement en fonction de l'algorithme. Enfin on juge de la \textbf{complexité à calculer une clé} pour chacune des machines du groupe, qui varie en fonction des primitives cryptographiques utilisées et de leur nombre (voir table \ref{tab:intro:tgdh_vs_str}). \ac{PFS}, qui assure qu'un adversaire ne peut déchiffrer les messages précédant la compromission d'une clé principale d'un membre ; couplé à une résistance aux adversaires pouvant écouter et modifier le traffic. En ce qui concerne l'authentification, nous assumons que la couche réseau ou applicative sous-jacente permet déjà l'authentification de l'auteur du contenu : ainsi pour \ac{ICN}, le \emph{producteur} de contenu fournirait un \ac{HMAC} signant les manifestes listant les paquets émis pour un contenu.

\vspace{-12pt}
\begin{table*}[h!]
	\centering
	\begin{tabular}{@{}lcccccccccc@{}}
		Protocole & \multicolumn{3}{c}{Tours} & \rot{Adversaire} & Autres Propriétés & \rot{Open Source} \\ 
		\tiny{référence (nom)} & \tiny{(init)} & \tiny{join} & \tiny{leave} & & & \\ \midrule
		
		\cite{ingemarsson_conference_1982} (ITW)      & & \cellcolor{black!15} $m-1$ & \cellcolor{black!15} $m-1$     & \cellcolor{black!15} P & \cellcolor{black!15} & \ding{51}  \\
		
		\cite{burmester_secure_1994} (BD) & & 2 & 2 & \cellcolor{black!15} P & \cellcolor{black!15} & \ding{51} \\
		
		\cite{kim_simple_2000} (STR) & & 2 & 1        & A & PFS & \ding{51} \\
		
		\cite{kim_simple_2000} (TGDH) & & 2 & 1& A & PFS & \ding{51} \\
		
		\cite{lee_efficient_2003} (P-TGDH) & & 2 & 1 & A & PFS & \cellcolor{black!15}\ding{55} \\
		
		\cite{zou_block-free_2004} (BF-TGDH) & & 2 & 1 & A & PFS & \cellcolor{black!15}\ding{55} \\
		
		\cite{striki_robust_2006} (DS-TGDH) & & 2 & 1 & A & PFS, optimisation d'arbre & \cellcolor{black!15}\ding{55} \\
		
		\cite{hietalahti_clustering-based_2008} (BD-AT-GDH) & & 2 & 2 & A & PFS, clusters & \cellcolor{black!15}\ding{55}\\
		
		\cite{zheng_communicationcomputation_2007} (CCEGK) & & 2 & 1 & A & PFS & \cellcolor{black!15}\ding{55} \\
		
		\cite{sunghyuck_queue-based_2009} (QGKA) & & 2 & 1 & A & PFS & \cellcolor{black!15}\ding{55} \\
		
		\bottomrule
		
	\end{tabular}
	
	\caption[Comparaison des protocoles d'établissement de clé de groupe]{Comparaison des protocoles d'établissement de clé de groupe (omettant des protocoles similaires à ceux non-retenus et parus plus récemment)\\ $m$: nombre de membre du groupe | Adversaire: \textbf{A}ctif/\textbf{P}assif | propriétés écartant un protocole : gris foncé}
	\label{tab:intro:new_proto_selection}
\end{table*}

Depuis que l'échange de clés Diffie-Hellman multipartite a été initialement proposé en 1976, de nombreux travaux on cherché à l'adapter au contexte d'un groupe. Ingemarsson et al. (in 1982) \cite{ingemarsson_conference_1982} and Burmester/Desmedt (BD algorithm) (1994) \cite{burmester_secure_1994} ont notamment proposé une solution simple - mais malheureusement impossible à mettre en pratique puisque : il assume que le réseau est capable de broadcast simultané là où en pratique il faut un tour pour chaque membre. La preuve générale du Diffie-Hellman de Groupe (GDH) prend cela en compte, mais ne s'encombre pas de l'authentification. Le premier protocole à synthétiser ces besoins est Tree Group Diffie-Hellman \cite{kim_simple_2000}. Suivirent quantité de propositions de protocoles se reposant sur l'architecture de TGDH, (dont les propriétés sont détaillées \tabref{intro:new_proto_selection}), et la plupart gardent un nombre similaire de tours si ce n'est dans des cas particuliers.

\begin{table*}[h!]
	\centering
	\label{tab:intro:tgdh_vs_str}
	\begin{tabular}{@{}lrcccc|ccc@{}}
		\multicolumn{1}{l}{Protocole} & & \multicolumn{4}{c}{Communication} & \multicolumn{3}{c}{Computation} \\ 
		\tiny{référence (nom)} & & \tiny{tours} & \tiny{messages} & \tiny{unicast} & \tiny{multicast} & \tiny{exponentiations} & \tiny{signatures} & \tiny{vérifications} \\
		
		\midrule
		
		\multirow{3}{*}{\cite{kim_simple_2000} (TGDH)}&
			\tiny{Join, Merge}	& 2 & 3 & 0 & 3 & $\frac{3h}{2}$ & 2 & 3 \\
		&	\tiny{Leave}    	& 1 & 1 & 0 & 1 & $\frac{3h}{2}$ & 1 & 1 \\
		&	\tiny{Partition}	& $h$ & $2h$ & 0 & $2h$ & $3h$ & $h$ & $h$ \\
		
		\midrule
		
		\multirow{3}{*}{\raggedleft{\cite{kim_simple_2000} (STR)}}&
			\tiny{Join}					& 2 & 3 & 0 & 3 & 7 & 2 & 3 \\
		&	\tiny{Leave, Partition}		& 1 & 1 & 0 & 1 & \cellcolor{black!15}$\frac{3n}{2} + 2$ & 1 & 1 \\
		&	\tiny{Merge}				& 2 & 3 & 0 & 3 & \cellcolor{black!15}$3m + 4$ & $2$ & $3$ \\
		
		\bottomrule
		
	\end{tabular}
	\caption[Comparaison des coûts communicationels/computationels pour TGDH et STR]{Comparaison des coûts communicationels/computationels pour TGDH et STR\\ $m$: nombre de membres du groupe, $n$: nombre de nœuds de l'arbre, $h$: hauteur de l'arbre | propriétés écartant un protocole : gris foncé}
\end{table*}

On peut arguer du fait que les améliorations apportées à TGDH peuvent être faites de manière incrémentale sur une implémentation existante de TGDH, constituant ainsi une base avec laquelle comparer l'amélioration proposée. Forts de ce constat, nous avons ainsi préféré sélectionner TGDH parmi les protocoles déjà sélectionnés dans la table \ref{tab:intro:new_proto_selection} et tout particulièrement STR qui était le plus prometteur mais ne croît pas logarithmiquement pour une opération \emph{leave}, \emph{partition} ou \emph{merge} (voir table \ref{tab:intro:tgdh_vs_str} ci-dessus).

\section{Implémentation de Tree Group Diffie-Hellman}

L'algorithme précédemment sélectionné, TGDH, fonctionne en organisant les membres et le calcul de leur clé selon un arbre binaire publique où chaque nœud partage un \emph{blinding} de sa clé privée. Le calcul de la clé du groupe est effectué par un membre particulier du groupe, appelé « sponsor ». À chaque ajout de membre, le sponsor réorganise l'arbre du groupe et recalcule les clés intermédiaires le long de la branche ayant été modifiée, avant de recalculer la clé du groupe et de partager l'arbre aux autres membres pour qu'ils n'aient que la dernière clé à recalculer. Des procédures correspondantes sont définies pour quitter, partitionner et fusionner des groupes \cite{kim_tree-based_2004}.

\begin{SCfigure}[][h!]
	\includegraphics[width=6cm]{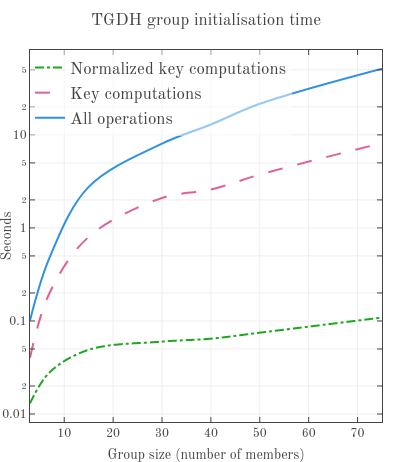}
	\caption[Temps total d'initialisation de TGDH, variant par taille de groupe]{Temps total d'initialisation de TGDH, variant par taille de groupe, lissé sur 100 itérations.\\Les calculs faits par le sponsor \emph{et} les autres membre pour vérifier et générer la clé sont pris en compte. Le temps total devrait être mis en perspective avec une implémentation na\"{i}ve de Group Diffie-Hellman, où $n-1$ calculs de clé sont nécessaires au lieu de $\mathcal{O}(\log_{2}{n})$ pour TGDH, et ce à chaque ajout de membre. Le programme de test présenté ici considère chaque membre comme sponsor pour pouvoir vérifier la validité du programme, aussi une normalisation par le nombre de membre s'impose pour comparer avec un cas d'usage pratique, où tous sauf le sponsor vérifient seulement la dernière clé et opèrent en parallèle.}
	\label{fig:init_time}
\end{SCfigure}

Afin de valider le bon fonctionnement de notre algorithme\footnote{le code source est disponible sur \href{https://framagit.org/rigelk/gka-libparc}{https://framagit.org/rigelk/gka-libparc}.} et des ses performances, nous avons simulé la création et l'ajout de plusieurs membres à un groupe sur un ordinateur de référence. Le temps d'exécution semble faible (voir \figref{init_time}) au regard de la procédure de test: passer de 0 à 70 membres se fait en 8.16s sur la machine de référence utilisée\footnote{un i7-4600U CPU @ 2.10GHz a été utilisé comme machine de référence pour tous les tests.}. La croissance polynomiale du temps d'initialisation peut être attribuée à deux facteurs : le nombre allant croissant avec le nombre de membres à considérer (un facteur confirmé par un profilage de l'exécution du programme) qui peut être évité avec une mise en cache correcte (une absence d'optimisation de notre implémentation et non un manque \emph{per se} de l'algorithme) ; le fait que le reste de l'algorithme de test soit écrit avec la volonté de rester simplement vérifiable et opère séquentiellement pour chaque membre re-calculant sa clé, une situation qui n'a pas lieu en pratique (seul le sponsor doit effecteur le calcul pour tous les membres, et tous peuvent vérifier uniquement la clé résultante en parallèle).

Le temps requis pour ajouter les membres précédents, une fois normalisé, est assez court pour être réaliste à l'échelle d'un groupe de partage de fichiers, et correspond déjà aux performances rapportées pour TGDH \cite{kim_tree-based_2004}. On note que le fait de grouper les opérations peut permettre une amélioration des performances en pratique, car le délai réseau ajouté entre chaque opération peut vite devenir significatif. Pour ce faire, il faut définir une temporisation afin d'agréger les ajouts de membres pour les traiter ensemble. Une autre catégorie d'optimisation de l'algorithme, plutôt réservée aux réseaux d'entreprise où l'on peut vouloir réduire le nombre de broadcasts et où l'on peut déployer une infrastructure dédiée, consiste à affecter une machine à la tache de stocker et servir les \emph{blinding}, à la manière d'un serveur de clé GPG. On diminue alors la réactivité du groupe en cas de partition, mais on gagne $n$ broadcasts. La même chose peut être faite pour désigner un sponsor fixe réservé au calcul et permettant d'éviter le calcul par des clients légers.

\section{Conclusion}

À travers cet article, nous avons étudié un effet de bord caractéristique du chiffrement concernant l'infrastructure de routage, et avons proposé une approche simple afin d'établir une clé de groupe de manière efficace dans un cas d'usage moderne : les réseaux centrés sur l'information. Cependant pour parvenir à une performance acceptable au-delà de la simple preuve de concept, des adaptations devront être apportées à des protocoles connus d'échange de clés comme celui sélectionné dans cet article. Gageons aussi que leur implémentation nouvelle permette de diffuser des techniques d'échange de clés de groupe jusqu'alors relativement confidentielles. La prochaine étape restera le développement des réseaux ICN et apparentés, mais de futurs travaux peuvent déjà s'orienter vers l'utilisation de clusters pour remplacer certaines branches relativement statiques de l'arbre d'un groupe (e.g. dans le cadre d'un réseau de machines fixes) tels que définis dans BD-AT-GDH \cite{hietalahti_clustering-based_2008} afin de diminuer la complexité à l'initialisation de certains réseaux.

\nocite{}
\bibliographystyle{alpha}
\bibliography{main}
\label{sec:biblio}

\end{document}